\documentclass[a4paper, conference, 11 pt]{IEEEtran}
\IEEEoverridecommandlockouts

\usepackage{algorithm}
\usepackage{algpseudocode}
\usepackage{pifont}

\usepackage{url}
\usepackage{cite}
\usepackage{amsmath,amssymb,amsfonts,bm}
\usepackage{graphicx}
\usepackage{textcomp}
\usepackage{xcolor}

\usepackage{color}
\usepackage{mathrsfs}

{ 

}
\usepackage{float}

\newcommand{\bi}{\begin{itemize}}
\newcommand{\ei}{\end{itemize}}
\newcommand{\bd}{\begin{displaymath}}
\newcommand{\ed}{\end{displaymath}}
\newcommand{\be}{\begin{eqnarray*}}
\newcommand{\ee}{\end{eqnarray*}}

\newcommand{\qed}{\nobreak \ifvmode \relax \else
      \ifdim\lastskip<1.5em \hskip-\lastskip
      \hskip1.5em plus0em minus0.5em \fi \nobreak
      \vrule height0.75em width0.5em depth0.25em\fi}

\begin{document}

\title{Model-Agnostic Algorithm for Real-Time Attack Identification in Power Grid using Koopman Modes\\ 
\thanks{Pacific Northwest National Laboratory (PNNL) is a multi-program national laboratory operated by Battelle for the U.S. Department of Energy (DOE) under contract No. DE-AC05-76RL01830. 
}
}


\author{\IEEEauthorblockN{Sai Pushpak Nandanoori\IEEEauthorrefmark{1},
Soumya Kundu\IEEEauthorrefmark{1},
Seemita Pal\IEEEauthorrefmark{2}, 
Khushbu Agarwal\IEEEauthorrefmark{3} and
Sutanay Choudhury\IEEEauthorrefmark{3}}
\IEEEauthorblockA{\IEEEauthorrefmark{1}Optimization and Control Group\\\IEEEauthorrefmark{2}Electricity Security Group\\ \IEEEauthorrefmark{3}Data Sciences Group\\
Pacific Northwest National Laboratory, USA\\
Emails:\{saipushpak.n,\,soumya.kundu,\,seemita.pal,\,khushbu.agarwal,\,sutanay.choudhury\}@pnnl.gov}}

\maketitle

\begin{abstract}
Malicious activities on measurements from sensors like Phasor Measurement Units (PMUs) can mislead the control center operator into taking wrong control actions resulting in disruption of operation, financial losses, and equipment damage. In particular, false data attacks initiated during power systems transients caused due to abrupt changes in load and generation can fool the conventional model-based detection methods relying on thresholds comparison to trigger an anomaly. In this paper, we propose a Koopman mode decomposition (KMD) based algorithm to detect and identify false data attacks in real-time. The Koopman modes (KMs) are capable of capturing the nonlinear modes of oscillation in the transient dynamics of the power networks and reveal the spatial embedding of both natural and anomalous modes of oscillations in the sensor measurements. The Koopman-based spatio-temporal nonlinear modal analysis is used to filter out the false data injected by an attacker. The performance of the algorithm is illustrated on the IEEE 68-bus test system using synthetic attack scenarios generated on GridSTAGE, a recently developed multivariate spatio-temporal data generation framework for simulation of adversarial scenarios in cyber-physical power systems. 
\end{abstract}

\begin{IEEEkeywords}
cyber-physical security; Koopman mode decomposition; false data injection; online attack identification.
\end{IEEEkeywords}

\section{Introduction}

\begin{figure}[thpb]
	\centering
	\includegraphics[width=0.7\columnwidth]{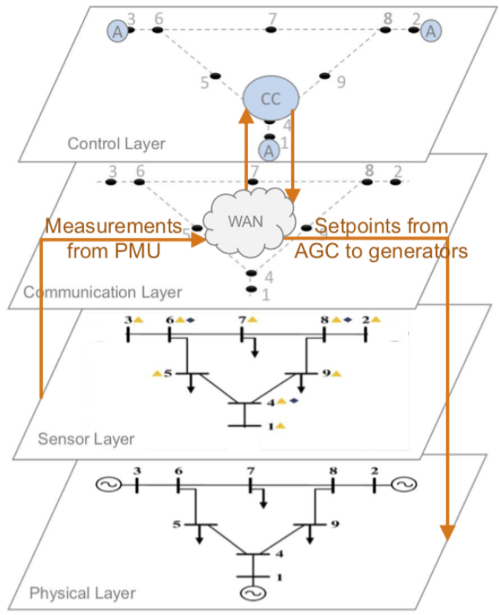}
	\caption{Illustration of a closed-loop power system, including the controller (e.g. AGC), sensors (e.g. PMUs), actuators (e.g. generators) and the communication layer (via WAN).}
	\label{fig:cps}
\end{figure}

Modern electrical grids are increasingly complex cyber-physical systems, with advanced sensing, control and communication layers overlaid on nonlinear dynamical networks.  Real-time grid operations are aided by closed-loop feedback control decisions that rely on advanced sensor measurements to dynamically secure the balance between supply and demand. For instance, the \textit{automatic generation control} (AGC), illustrated in Fig.\,\ref{fig:cps}, is designed to process measurements from advanced sensors such as \textit{phasor measurement units} (PMUs) to update the governor set-points of the generators every few seconds. Such tightly coupled cyber-physical control operation arguably presents an opportunity for malicious cyber agents to sneak into the system and cause disruptions, equipment damages and/or financial losses. Real-world events such as the \textit{Stuxnet} \cite{karnouskos2011stuxnet}, the \textit{Dragonfly} \cite{team2017dragonfly}, and the 2015 cyber-attacks on the Ukrainian power grid \cite{case2016analysis} and on a substation in the state of California in US \cite{reilly2015bracing}, demonstrate the vulnerability of power systems worldwide to cyber attacks. Different, possibly overlapping, classes of cyber-attacks have been investigated in the power systems community, including \textit{denial of service} (DoS) attacks \cite{pushpak2014vulnerability,kurt2018distributed}; \textit{gray-hole} or packet drop attacks \cite{Pal2018AnOM}; \textit{jamming} or link-failure attacks  \cite{kurt2018real,pushpak2014vulnerability}; and \textit{false data injection} (FDI) or \textit{data integrity} attacks \cite{bobba2010detecting,liu2011false,xie2011integrity,pushpak2014vulnerability,kurt2018distributed}. 

\textbf{Related Work} Due to economic and other reasons, sensors are arguably among the least-protected components in a power network \cite{ferragut2017real} and, therefore, particularly vulnerable to malicious cyber attacks. As such, we focus in this work on identification of FDI attacks on PMUs which can report GPS time-stamped measurements at typically 20-60 samples per second \cite{Pal2018AnOM} and allows visibility into the system transient dynamics. A broad category of research efforts that considers power systems transient dynamics while detecting and identifying sensor attacks relies on physics-based models, coupled with a dynamic state estimator (e.g. Kalman filters), to detect anomalous sensors data via outlier detection methods such as \textit{cosine similarity test} or statistical \textit{$\chi^2$-test} \cite{manandhar2014detection,rawat2015detection,ghosal2018diagnosis}. In \cite{ghosal2018diagnosis}, a state-space transformation and decomposition technique was proposed to distinguish between the anomalies caused due to malicious attacks and physical changes (e.g. changes in topology, load, generation). In \cite{murguia2016characterization}, the authors extended the conventional \textit{static} outlier detection method of evaluating one sample of measurements at a time to a \textit{dynamic} change-point detector based on the \textit{cumulative sum} (CUSUM) test on a moving window of measurements. The work in \cite{huang2018online} presents an attack detection algorithm that involves superimposing \textit{watermarked} signals on the control inputs transmitted over secure channels. In addition to the physics-based methods mentioned above, machine learning-based online attack detection algorithms have been proposed in \cite{chen2016multi,kurt2018online}. 

\textbf{Approach} Most of the existing detection methods are often unreliable during system transients due to unforeseen changes or insufficient training data. In addition, the system dynamics is expected to display rich nonlinear behavior during the transients. In this work, we propose an online attack detection algorithm which uses the Koopman mode decomposition (KMD) to identify the spatial embedding of anomalous (and natural) modes in the sensors data. 
%
The \textit{Koopman operator} is a linear, infinite-dimensional, composition operator that describes the temporal evolution of a set of observable functions along the trajectories of a finite-dimensional (nonlinear) dynamical system \cite{koopman1931hamiltonian,susuki2016applied}. Spectral analysis of the Koopman operator reveals that single-frequency modes, henceforth termed as the \textit{Koopman modes} (KMs), can be embedded in the spatio-temporal dynamics of the nonlinear systems \cite{rowley2009spectral}. The KMs are particularly promising in power systems applications, by way of extending the notions of linear modal analysis (e.g. participation factors) to the nonlinear regime in a computationally efficient manner in contrast to other methods such as normal forms \cite{netto2018data,sanchez2005inclusion}.
Efficient algorithms exist to compute KMs, as well as the Koopman operator from streaming data in power systems \cite{barocio2014dynamic,sinha2020data}. Koopman modal analysis has found several applications in power systems, including coherency detection, stability, and partitioning (see \cite{susuki2016applied} and the references therein for details). As a novel contribution of this work, we extend the Koopman modal analysis to the case of real-time identification (detection and localization) of malicious data attacks on power system sensors, in presence of transient fluctuations. 

\textbf{Contributions} In this paper, we develop an online attack identification algorithm that performs a Koopman-based spatio-temporal nonlinear modal analysis on streaming data from the sensors to detect and localize malicious activities. Performance of the algorithm is illustrated on IEEE 68-bus test-case using synthetic attack data generated using GridSTAGE \cite{gridstage2020} which is a simulations-based framework (built upon \cite{zhang2016augmenting} and \cite{chow1992toolbox}) to synthesize adversarial scenarios in a cyber-physical power grid. The rest of the article is structured as follows: in Sec.\,\ref{S:problem}, we present the attack identification problem in a closed-loop power systems; Sec.\,\ref{S:background} briefly recaps the KMD, while the proposed attack identification algorithm is described in Sec.\,\ref{S:method}; Sec.\,\ref{S:results} illustrates the working of the algorithm with numerical simulation results; before the article is concluded in Sec.\,\ref{S:concl} with a summary and future plan.


\section{Problem Description}\label{S:problem}

The power grid is a nonlinear electro-mechanical dynamical system primarily driven by the complex interactions of various synchronous generator models, excitation systems, power system stabilizers, governor controls, as well as dynamic load modulation \cite{chow1992toolbox,zhang2016augmenting}. Since the attack identification algorithm does not require any knowledge of the system model, it suffices to consider only a compact abstraction of the system dynamics in the control-affine form of:
\begin{subequations}\label{E:sys}
\begin{align}
    x_{k+1} &= f(x_k) + h(x_k)\,u_k \\
    y_k &= g(x_k)
\end{align}\end{subequations}
where $x_k$ are the vector-valued internal states (e.g. voltage angles, voltage magnitudes, and frequencies); $y_k$ are the vector-valued measurements from sensors (e.g. PMUs); $u_k$ are the vector-valued control inputs to the generators (e.g. governor set-points); $f$, $g$ and $h$ are locally Lipschitz vector-valued functions of states representing the system dynamics, measurements, and control channels respectively. For notational convenience, we assume that the nominal (target) operating point is shifted to the origin (i.e. $f(0)\!=\!0$), and that $g(0)\!=\!0$\,.

\begin{figure}[htpb]
	\centering
	\includegraphics[scale=0.15]{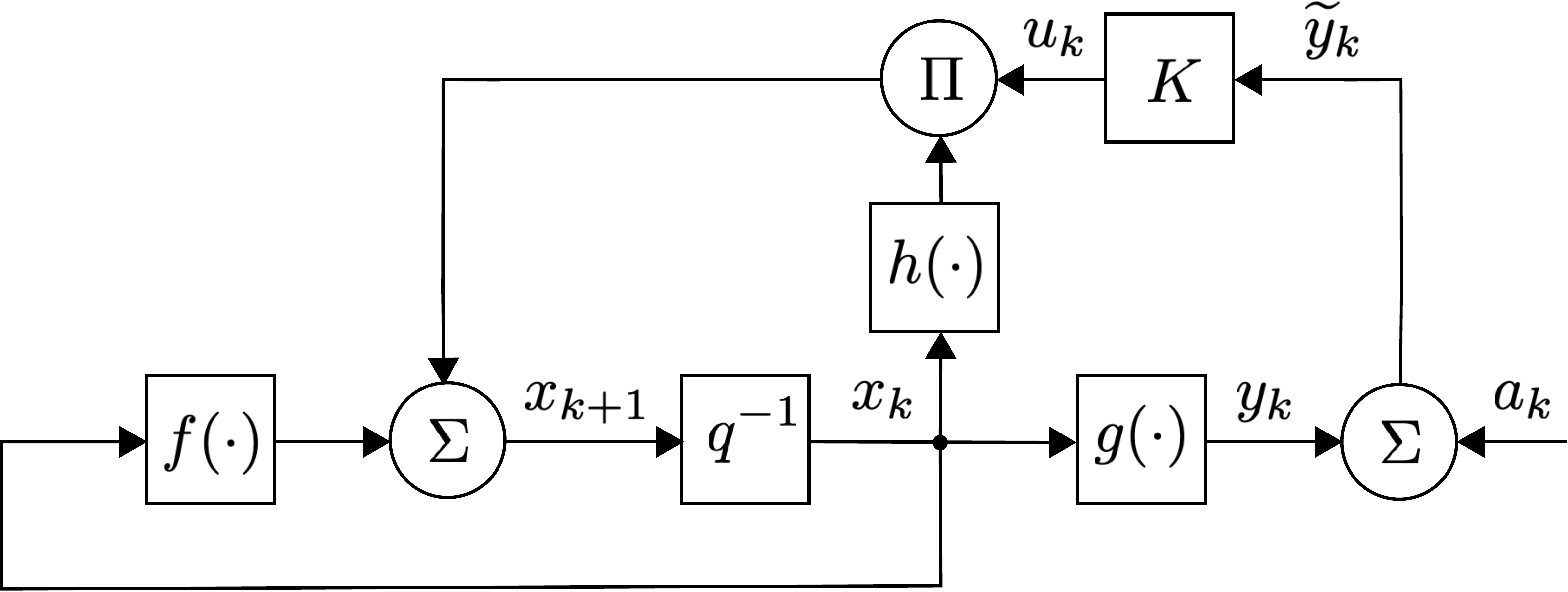}
	\caption{Block diagram of the closed-loop system under attack.}
	\label{fig:closed_loop}
\end{figure}

\textbf{Attack Model:} The adversary is assumed to have compromised one or more of the PMUs, \textit{phasor data concentrators}, network routers and/or communication links. Note that since IEEE Standard C37.118-2 places no restriction on the choice of the communication medium, portions of the PMU communication network may include the internet which is vulnerable to cyber-attacks. At each of the compromised nodes in the communication network, the adversary can manipulate PMU data which is typically not encrypted. The control center receives the manipulated measurements data $\widetilde{y}_k$, with $a_k\!:=\!\widetilde{y}_k\!-\!y_k$ being the vector-valued \textit{injected attack} signal. The control inputs are computed using a proportional output-feedback controller with control gain $K$, such that the closed-loop system becomes:
\begin{subequations}\label{E:sys_con}
\begin{align}
    x_{k+1} &= f_c(x_k) + h_c(x_k)\,a_k\\
    \widetilde{y}_k &= g(x_k)+a_k
\end{align}
\end{subequations}
where $h_c(x_k)\!:=\!h(x_k)\,K$ and $f_c(x_k)\!:=\!f(x_k)\!+\!h_c(x_k)\,g(x_k)$. 
Fig.\,\ref{fig:closed_loop} illustrates the closed-loop system including the control inputs and the attack signals. 
Note that we do not place any restriction on the type of attack, and merely interpret the attack signal ($a_k$) as any maliciously introduced anomaly between the actual measurements ($y_k$) and the measurements received at the control center ($\widetilde{y}_k$).

\textbf{Attack Identification:} Attack identification algorithms involve construction of vector-valued \textit{measurement residuals} ($r_k$) as a difference between the measurements received at the control center ($\widetilde{y}_k$) and the estimated measurements ($\widehat{y}_k$):
\begin{align}\label{E:residual}
    r_k:=\widetilde{y}_k-\widehat{y}_k\,.
\end{align}
The measurement residuals are fed into some anomaly detector, such as \textit{cosine similarity} test, $\chi^2$-test, or CUSUM test \cite{manandhar2014detection,rawat2015detection,ghosal2018diagnosis,murguia2016characterization,huang2018online}. However, the estimates of the measurements are often inaccurate (or, unavailable) during power system transients caused by unforeseen and/or unknown changes in load and generation - potentially misleading the attack detectors into \textit{false positives} (natural events identified as attacks) or \textit{false negatives} (labeling an attack event as benign). While the work in \cite{ghosal2018diagnosis} proposes a method to distinguish between anomalies due to attacks and modeling errors, its reliance on physics-based linear system models and determination of multiple thresholds for statistical tests potentially limits its application to large-scale power systems displaying rich nonlinearity. 

In this work, we propose a Koopman-based attack identification algorithm which uses a moving window of $n\!+\!1$ most recent snapshots of measurements $\lbrace \widetilde{y}_{k\!-\!n}, \dots,\widetilde{y}_{k\!-\!2}, \widetilde{y}_{k\!-\!1}, \widetilde{y}_k\rbrace$, $n\geq2$, to identify in real-time the sensors under attack. 

\section{Background: Koopman Mode Decomposition}\label{S:background}

Consider the discrete-time closed-loop system \eqref{E:sys_con} under the scenario when there is no attack, i.e. $a_k\!=\!0\,\forall k$, 
\begin{align}\label{E:sys_no_attack}
    \text{(no attack)}\quad x_{k+1}&=f_c(x_k)
\end{align}
The \textit{Koopman operator} $\mathcal{U}$, \cite{susuki2016applied,netto2018data}, is an infinite-dimensional linear operator that acts on any scalar-valued function of the state, $\psi$, and maps it into another scalar-valued function of the state $\mathcal{U}\psi$ defined as follows:
\begin{align}\label{E:koopman}
    \mathcal{U}\psi := \psi\circ f_c\,.
\end{align}
The linearity of the operator is exemplified by the fact that for any two scalars $\alpha_1$ and $\alpha_2$\,, and any two scalar-valued functions of state $\psi_1$ and $\psi_2$, we have
\begin{align*}
    \mathcal{U}\left(\alpha_1\psi_1+\alpha_2\psi_2\right)&=\alpha_1\left(\mathcal{U}\psi_1\right) + \alpha_2\left(\mathcal{U}\psi_2\right)
\end{align*}
The Koopman operator $\mathcal{U}$ admits (infinitely many) Koopman eigenfunctions $\phi_j$ (non-zero scalar-valued functions of state) and Koopman eigenvalues $\lambda_j\!\in\!\mathbb{C}$ defined as follows:
\begin{align}\label{E:eigen}
    \mathcal{U}\phi_j=\lambda_j\phi_j\,,\quad j=1,2,3,\dots
\end{align}
Note that iterative application of \eqref{E:koopman} and \eqref{E:eigen} yields
\begin{align}
    \phi_j(x_k)=\mathcal{U}^k\phi_j(x_0)=\lambda_j^k\phi_j(x_0)\,.
\end{align}
The set of all Koopman eigenvalues ($\lambda_j$) forms the \textit{discrete spectrum} of the Koopman operator $\mathcal{U}$ and, along with its continuous spectra, define the spectral properties of $\mathcal{U}$ \cite{susuki2016applied}. 

\textbf{Koopman Mode Decomposition:} The Koopman operator translates the finite-dimensional nonlinear dynamics into an infinite-dimensional linear representation. In particular, let us consider the measurement function $g=\left(g_1,\,g_2,\,\dots,\,g_p\right)^T$ as a $p$-dimensional vector-valued function of observables. If each $g_i$ lies within the span of the Koopman eigenfunctions $\phi_j$, then the vector-valued observables $g$ can be expanded as:
\begin{align}\label{E:KMD}
    g(x_k) = {\sum}_{j=1}^\infty\phi_j(x_k)v_j={\sum}_{j=1}^\infty\lambda_j^k\phi_j(x_0)v_j
\end{align}
where $v_j$ are $p$-dimensional vector-valued coefficients of the expansion, called the \textit{Koopman modes} (KMs) \cite{susuki2016applied}. The expansion in \eqref{E:KMD} is referred to as the \textit{Koopman Mode Decomposition} (KMD). Moreover, $\lambda_j$ encode the temporal signatures in the dynamics, with associated spatial signatures encoded in $\phi_j(x_0)v_j$\,. In particular, the magnitude and phase of the Koopman eigenvalue $\lambda_j$ are called, respectively, the \textit{growth rate} and the \textit{frequency} of the associated KM. 

\textbf{Empirical Computation:} There exist efficient algorithms to compute finite-sum approximations of the KMD \eqref{E:KMD} based on, for example, the Arnoldi algorithm \cite{rowley2009spectral,susuki2016applied} and the extended dynamic mode decomposition \cite{klus2015numerical,netto2018data}. While we avoid a description of the methods here for space constraint, it is sufficient to note that the algorithms use a sequence of $n\!+\!1$ vector-valued observations 
$\left\lbrace g(x_0),g(x_1),\dots,g(x_{n-1}),g(x_n)\right\rbrace$, $n\!\geq\!2$, to obtain KMD as:
\begin{align}\label{E:KMD_approx}
     g(x_k) \approx {\sum}_{j=1}^n\widehat{\lambda}_j^k\,\widehat{v}_j
\end{align}
where $\widehat{\lambda}_j$ are estimates of the Koopman eigenvalues ($\lambda_j$)\,, while $\widehat{v}_j$ estimate the products of the Koopman eigenfunctions ($\phi_j(x_0)$) and the KMs ($v_j$). A finite dimensional approximation of the Koopman operator $K\!\!\in\!\mathbb{R}^{p\times p}$ is given as
\begin{align}\label{E:KO_approx}
    K = K_1\,K_2^+,\quad&\text{where}\\
K_1\!=\!\frac{1}{n}\sum_{k=0}^{n-1} g(x_{k+1})g(x_k)^T\!&\text{ and }K_2\!=\!\frac{1}{n}\sum_{k=0}^{n-1} g(x_k)g(x_k)^T\!,\notag
\end{align}
and the superscript `$^+$' denotes the \textit{pseudo-inverse}, \cite{klus2015numerical}. Notice that the empirical estimate $K$ of Koopman operator, learnt using $\left\lbrace g(x_0),\dots,g(x_n)\right\rbrace$ can be used to predict future observations $\left\lbrace \widehat{g}(x_{n+1}),\widehat{g}(x_{n+2}),\widehat{g}(x_{n+3}),\dots\right\rbrace$ as
\begin{align}\label{E:KO_predict}
    \widehat{g}(x_{n+m})=K^m\,g(x_n)\quad\forall m=1,2,3,\dots
\end{align}
The empirical computational steps \eqref{E:KMD_approx}-\eqref{E:KO_predict} will be used to design the attack identification algorithm in Sec.\,\ref{S:method}.

\section{Proposed Attack Identification Algorithm}\label{S:method}

\begin{figure*}
	\centering
	\vspace{0.3in}
	\includegraphics[width= 0.9 \textwidth]{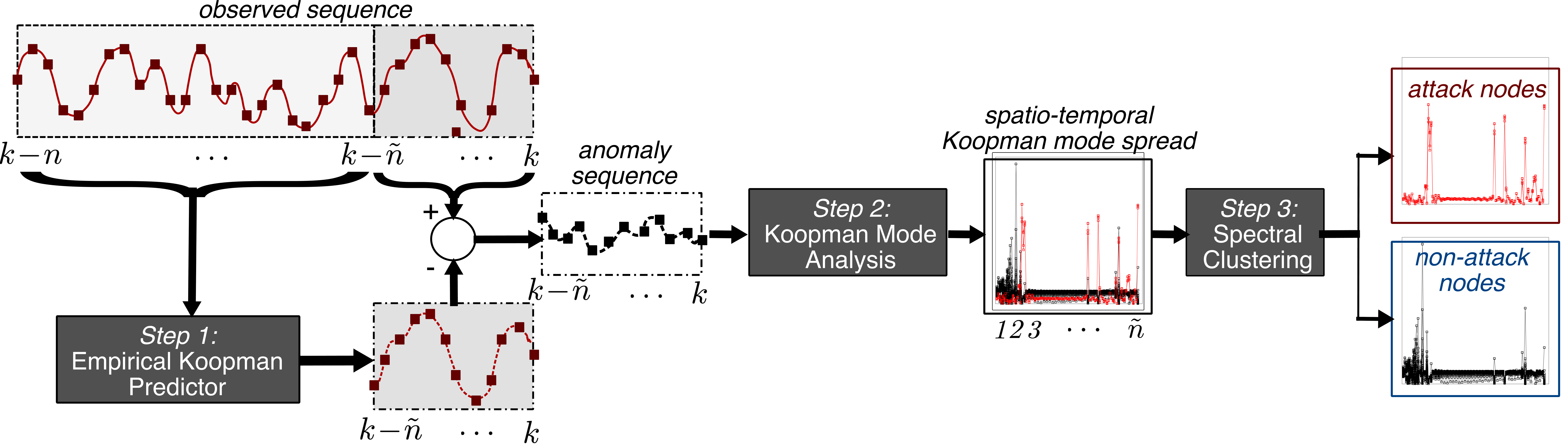}
    \vspace{-0.1in}
	\caption{Step-wise schematic of the proposed online attack identification algorithm.}
	\vspace{-0.1in}
	\label{fig:algorithm}
\end{figure*}

Fig.\,\ref{fig:algorithm} depicts key steps involved in the proposed online attack identification algorithm combining KMD with spectral clustering, as explained below. For identifying an attack at time $k$, a moving window sequence of $n\!+\!1$ observations are used, which is split into two sub-sequences: a (longer) \textit{learning window} $\mathcal{L}\!\in\!\lbrace k\!-\!n,\dots, k\!-\!\widetilde{n}\!-\!1\rbrace$, and a (shorter) \textit{prediction window} $\mathcal{P}\!\in\!\lbrace k\!-\!\widetilde{n},\dots,k\rbrace$\,, for some $\widetilde{n}\!<\!n$\,.

\textbf{Step\,1 Prediction:} We identify the finite-dimensional empirical Koopman operator $K$ on the \textit{learning window} $\mathcal{L}$ from \eqref{E:KO_approx} and using the empirical Koopman operator to predict the observations ($\widehat{g}$) within the \textit{prediction window} $\mathcal{P}$ from \eqref{E:KO_predict}. 

\textbf{Step\,2 Koopman Mode Analysis:} We compute the \textit{error} sequence ($\widetilde{g}$) by subtracting the predicted sequence ($\widehat{g}$) from the actually observed sequence during the \textit{prediction window} $\mathcal{P}$. Next the empirical KMs $\widehat{v}_j$ are computed using \eqref{E:KMD_approx} from the \textit{error sequence} ($\widetilde{g}$). 

\begin{algorithm}
\caption{Koopman-Based Online Attack Identification}
\label{ALG:steps}
\begin{algorithmic}[]
\State Choose $n$
\State Choose $\tilde{n}<n$
\State Set $k\geq n$
\While{\textit{true}}
\State Set \textit{learning window} $\mathcal{L}\!=\!\lbrace k\!-\!n,\dots,k\!-\!\tilde{n}\!-\!1\rbrace$
\State Set \textit{prediction window} $\mathcal{P}\!=\!\lbrace k\!-\!\tilde{n},\dots,k\rbrace$
\Procedure{Step 1 Prediction}{}
\State Estimate $K$ from \eqref{E:KO_approx} using the observation sequence $\lbrace g(x_l) \rbrace_{l\in\mathcal{L}}$ (i.e. the \textit{learning window)} 
\State Use $K$ to predict, as per \eqref{E:KO_predict}, the observation sequence $\lbrace \widehat{g}(x_l) \rbrace_{l\in\mathcal{P}}$ (i.e. the \textit{prediction window})
\EndProcedure
\Procedure{Step 2 Koopman Mode Analysis}{}
\State Compute error sequence $\widetilde{g}(x_l)\!=\!g(x_l)\!-\!\widehat{g}(x_l)$ for every $l\!\in\!\mathcal{P}$ (i.e. the \textit{prediction window})
\State Compute the empirical KMs from the error sequence $\lbrace \widehat{g}(x_l) \rbrace_{l\in\mathcal{P}}$ using \eqref{E:KMD_approx} 
\EndProcedure
\Procedure{Step 3 Spectral Clustering}{}
\State Spatio-temporally normalize KMs
\State Perform spectral clustering on the normalized KMs by using the weighted adjacency matrix
\EndProcedure
\State $k \gets k+1$
\EndWhile
\end{algorithmic}
\end{algorithm}

\textbf{Step\,3 Spectral Clustering} The KMs identified in Step\,2 encode the latent spatio-temporal signatures within the error sequence required to identify the adversarial attack. The intuitive idea is that if there is an attack during the \textit{prediction window} $\mathcal{P}$ then it should leave certain anomalous signature within the KMs specific to the attacked sensor measurements. On the other hand, if there is a natural event (sudden load or generation change), that would excite similar modes in the physically neighboring nodes reflecting as non-anomalous deviations within the KMs. This allows differentiating between adversarial attacks and natural events. To facilitate the identification of anomalous sensor measurements, we perform a spatio-temporal normalization on the KMs $\widehat{v}_j$\,. First, we stack up the absolute values of the KMs into a $p\!\times\!\widetilde{n}$ matrix:
\begin{align}
    \bm{\widehat{v}}:=\left[\left|\widehat{v}_1\right|~\left|\widehat{v}_2\right|~\dots~\left|\widehat{v}_{\widetilde{n}}\right|\right]
\end{align}
Then each column of $\bm{\widehat{v}}$ is normalized to unity-sum, which effectively puts \textit{equal} weight to each KM or, equivalently, disregards the impact of the associated Koopman eigenvalues. Next, each row in the resulting matrix is normalized to unity-sum, which effectively puts \textit{equal} weight to each sensor irrespective of whether it is from a generator or a load. This results in the spatio-temporally normalized KMs stacked up into $\bm{\overline{v}}$\,. Each row in $\bm{\overline{v}}$ sums up to unity, and represent a sensor.

Finally a spectral clustering is performed to identify the anomalous sensors from the spatio-temporally normalized KMs. Since each row in $\bm{\overline{v}}$, corresponding to a sensor, can be seen as a \textit{probability mass function}, we use the Kullback–Leibler divergence between the (sensor) rows to construct the weighted adjacency matrix which is then fed to \textit{k-means clustering algorithm} \cite{ng2002spectral} to identify the dominant (attacked and non-attacked) clusters in the normalized KMs. 

The key algorithmic steps are summarized in Algorithm\,\ref{ALG:steps}.


%
%


\section{Numerical Results}\label{S:results} 

We begin this section by describing the adversarial data generation framework GridSTAGE, and then present an adversarial use-case illustrating the application of proposed Koopman-based attack identification algorithm. 

\textbf{GridSTAGE} \textit{(Spatio-Temporal Adversarial scenario GEneration)} To evaluate the performance of the closed-loop power system under several attack strategies and mitigation efforts, we developed \textit{GridSTAGE} which is a multivariate spatio-temporal data generation framework for simulation of adversarial scenarios in cyber-physical systems \cite{gridstage2020}. There is no existing adversarial data-generation framework that can incorporate several attack characteristics and yield adversarial PMU data.  GridSTAGE is available as open source from GitHub.   


GridSTAGE models the cyber-physical system of the power grid, simulates adversarial scenarios in the system and generates multi-variate, spatio-temporal network data. GridSTAGE is developed based on Power System Toolbox (PST) on which nonlinear time-domain simulations can be generated for standard IEEE bus systems. Using GridSTAGE, one can create several event scenarios by enabling or disabling any of the following: faults, AGC control, PSS control, exciter control, load changes, and different types of cyber-attacks. IEEE bus system data is used for defining the power system environment. Sensors in the power system include both PMU and SCADA, and simulated data are generated for both type of sensors. Rate of frequency and location of the sensors can be adjusted as well. Detailed instructions on generating several scenarios with different attack characteristics, load characteristics, sensor configuration, control parameters can be found in the github repository \cite{gridstage2020}. 
%
The GridSTAGE framework currently supports simulation of \textit{FDI} attacks (such as ramp, step, random, trapezoidal, multiplicative, replay) and \textit{DoS} attacks (such as time-delay, packet-loss, freezing) on PMU data \cite{huang2009understanding}. We use one of these attacks to demonstrate the proposed attack identification algorithm, while more extensive analysis on different types of attacks will be performed in future.  
%

\begin{figure}[thpb]
	\centering
	\vspace{-0.2in}
	\includegraphics[width=0.5\textwidth]{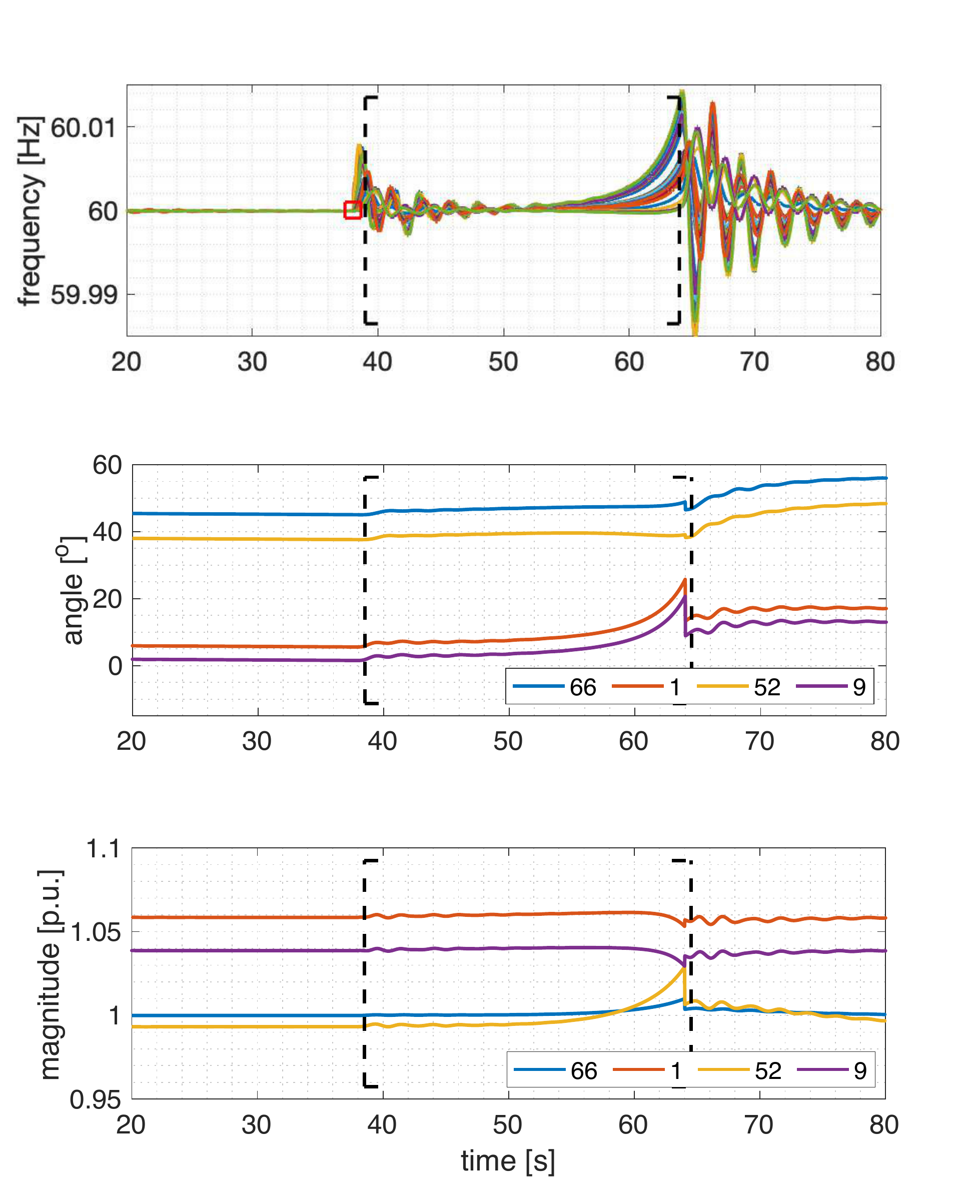}
	\caption{Time-series closed-loop simulation (using GridSTAGE) of an adversarial attack scenario. Bottom two plots show the \textit{attacked} PMU measurements of voltage angle and magnitude at the buses 1, 9, 52 and 66 in IEEE 68-bus network. Top plot shows the (delayed) impact of the attack on the resulting system frequencies. The attack duration (shown in black dashed lines) are between 39s and 64s, while a natural load change happens in bus 23 (marked by red square) at 38s.}
	\vspace{-0.2in}
	\label{fig:sim}
\end{figure}

\begin{figure}[thpb]
	\centering
	\includegraphics[width=0.5\textwidth]{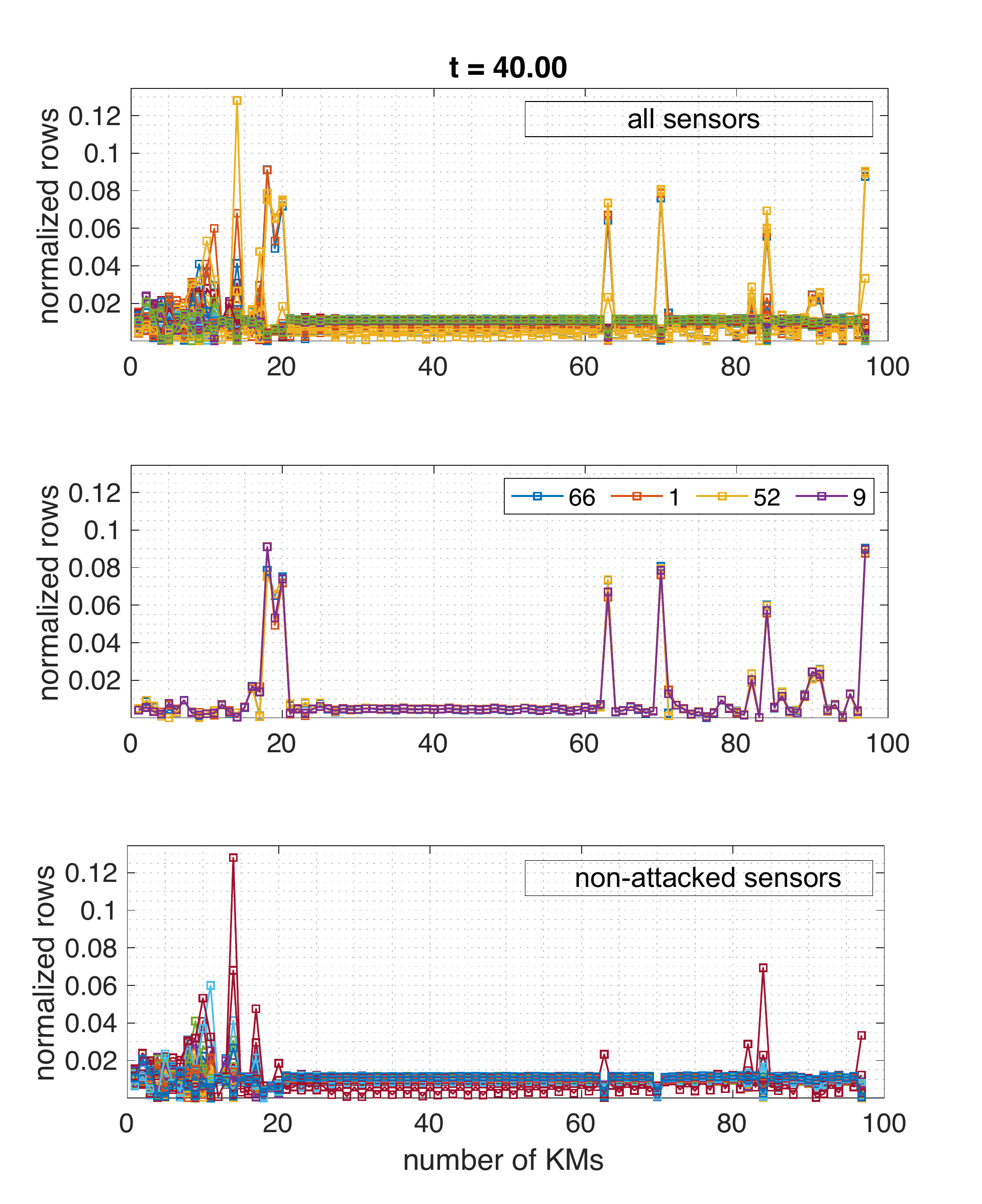}
	\caption{Koopman-based modal analysis followed by spectral clustering separates out the attacked sensors from the non-attacked one, even during transients. The attack is correctly identified at 40s, within 1s of the attack initiation.}
	\label{fig:modes}
\end{figure}

\textbf{Identifying Multiplicative Attack on PMUs}: In this subsection, we briefly discuss a multiplicative attack on the PMU data using the IEEE 68-bus as example. In this attack scenario, the attacker strategically initiates an FDI attack \textit{soon after} there is a load change in the system, by injecting a signal that grows over time in proportion to the disturbance being measured at the sensor. The attacker's goal is to \textit{hide behind} the existing transient disturbances in the network (caused due to load change). Based on the attack characteristics, this introduces a delayed impact on the system measurements and especially when the attack is over, the system sees a sudden change in measurements which engages control resources to counter the disturbance. Ideally, these resources should not have been engaged. The motivation of this work is to identify such attacks in their early stage such that disturbance seen by the system is deemed as attack and additional resources are not engaged. In Fig.\,\ref{fig:sim}, the attack scenario is illustrated, which is initiated at 39s right after a natural load change which happens at 38s. Finally Fig.\,\ref{fig:modes} illustrates a successful application of the online attack identification algorithm which is able to correctly identify all the attack locations (buses 1, 9, 52 and 66)) within 1s of the attack initiation, by performing a spectral clustering on the normalized Koopman mode spreads of all the sensors (top plot). It is evident that the Koopman mode spread for the attacked sensors (middle plot) are distinctively different from those at the other locations (bottom plot), and hence successfully picked out by the clustering algorithm.

\section{Conclusion}\label{S:concl}
This work introduced a novel real-time data-driven attack detection algorithm based on the Koopman mode decomposition and spectral clustering. Koopman modes embed the time-series streaming data (such as from PMUs) from the power grid network into spatio-temporal modes. A \textit{k-means} based spectral clustering on the spatio-temporally normalized KMs identifies the attacked and non-attacked clusters of sensors. The proposed attack detection algorithm is illustrated on a multiplicative attack on the PMU measurements representing the bus voltage angle and magnitudes generated using the GridSTAGE, a spatio-temporal multivariate adversarial data generation platform that is developed in the scope of this work. This induced attack if allowed to propagate for a few seconds, introduces unintended frequency changes in the network and engages resources without the need. However, the proposed algorithm successfully detects the induced attack within 1 second of attack initiation in the presence of load changes in the network. Future scope of this work will explore the application of the proposed attack detection algorithm on wider attack scenarios, including various strategic and stealthy attack types, as well as applicability of the method in differentiating between naturally caused events such as faults and component failures. 

\section*{Acknowledgment}
The authors thank Jiangmeng Zhang and Prof. Alejandro D. Domínguez-García at UIUC for providing their AGC code. 




\end{document}